\documentclass[conference,fleqn, letter, 10pt]{IEEEtran}

\usepackage[table]{xcolor}  
\usepackage{url}
\usepackage{mathtools}
\usepackage{multirow}
\usepackage{caption}
\usepackage{tcolorbox}
\usepackage[font=small]{subcaption}
\usepackage{enumitem}
\usepackage[
    bibstyle=ieee,
    citestyle=numeric-comp,
    maxnames=1,
    minnames=1
]{biblatex}
\AtBeginBibliography{\small}
\usepackage{lipsum}
\usepackage{stfloats}

\usepackage{acronym} \acrodef{6G}[6G]{sixth generation}
\acrodef{QoS}[QoS]{quality of service}
\acrodef{RIC}[RIC]{RAN intelligent controller}
\acrodef{DRL}[DRL]{deep reinforcement learning}
\acrodef{SMO}[SMO]{service management and orchestration}
\acrodef{Non-RT RIC}[Non-RT RIC]{non-real-time RIC}
\acrodef{Near-RT RIC}[Near-RT RIC]{near-real-time RIC}
\acrodef{VR}[VR]{virtual reality}
\acrodef{HMD}[HMD]{head-mounted display}
\acrodef{VR-CG}[VR-CG]{VR-cloud gaming}
\acrodef{3GPP}[3GPP]{3rd generation partnership project}
\acrodef{MEC}[MEC]{multi-access edge computing}
\acrodef{vRAN-UDP}[vRAN-UDP]{vRAN-unconstrained dynamic placement}
\acrodef{QoE}[QoE]{quality of experience}
\acrodef{AI}[AI]{artificial intelligence}
\acrodef{BS}[BS]{base station}
\acrodef{BS}[BS]{base station}
\acrodef{CN}[CN]{core network}
\acrodef{PN}[PN]{processing node}
\acrodef{NFV}[NFV]{network function virtualization}
\acrodef{SDN}[SDN]{software-defined networking}
\acrodef{RAN}[RAN]{radio access network}
\acrodef{vRAN}[vRAN]{virtualized \ac{RAN}}
\acrodef{MNO}[MNO]{mobile network operator}
\acrodef{MILP}[MILP]{mixed-integer linear programming}
\acrodef{MINLP}[MINLP]{mixed-integer nonlinear programming}
\acrodef{CU}[CU]{centralized unit}
\acrodef{DU}[DU]{distributed unit}
\acrodef{vCU}[CU]{virtualized central unit}
\acrodef{vDU}[DU]{virtualized distributed unit}
\acrodef{RU}[RU]{radio unit}
\newacroindefinite{RU}{an}{a}
\acrodef{O-RAN}[O-RAN]{open \ac{RAN}}
\acrodef{ILP}[ILP]{integer linear program}
\acrodef{TTI}[TTI]{transmission time interval}
\acrodef{URLLC}[URLLC]{ultra-reliable low-latency communications}
\acrodef{6G}[6G]{$6^{th}$ generation of wireless networks}
\acrodef{RIS}[RIS]{reconfigurable intelligent surfaces}
\acrodef{URLLC}[URLLC]{ultra-reliable low latency communication}
\acrodef{HRLLC}[HRLLC]{hyper-reliable low latency communication}
\acrodef{AI}[AI]{artificial intelligence}
\acrodef{ML}[ML]{machine learning}
\acrodef{DL}[DL]{deep learning}
\acrodef{BS}[BS]{base station}
\acrodef{SINR}[SINR]{signal-to-interference-and-noise ratio}
\acrodef{SNR}[SNR]{signal-to-noise ratio}
\acrodef{UE}[UE]{user equipment}
\acrodef{MD}[MD]{meta distribution}
\acrodef{3GPP}[3GPP]{3rd generation partnership project}
\acrodef{InF}[InF]{indoor factory}
\acrodef{UMa}[UMa]{urban macro}
\acrodef{DH}[DH]{dense high}
\acrodef{DL}[DL]{dense low}
\acrodef{SH}[SH]{sparse high}
\acrodef{SL}[SL]{sparse low}
\acrodef{LoS}[LoS]{line-of-sight}
\acrodef{NLoS}[NLoS]{non-line-of-sight}
\acrodef{ISD}[ISD]{intersite distance}
\acrodef{6G}[6G]{sixth-generation}
\acrodef{QoS}[QoS]{quality-of-service}
\acrodef{eCDF}[eCDF]{empirical cumulative distribution function}
\acrodef{CDF}[CDF]{cumulative distribution function}
\acrodef{eMBB}[eMBB]{enhanced mobile broadband}

\usepackage{geometry}
\usepackage{amsmath}
\usepackage{amssymb}
\usepackage{amsfonts}
\usepackage[ruled, linesnumbered]{algorithm2e}
\usepackage{algpseudocode}
\DeclareMathOperator*{\minimize}{minimize}

\iftrue
\usepackage{titlesec}

\titlespacing*{\section}{0pt}{\parskip}{-\parskip}
\titlespacing{\subsection}{0pt}{\parskip}{-\parskip}
\titlespacing{\subsubsection}{1em}{\parskip}{-\parskip}

\AtBeginDocument{
  \addtolength{\abovedisplayskip}{-4pt}
  \addtolength{\belowdisplayskip}{-4pt}
  \addtolength{\abovedisplayshortskip}{-2pt}
  \addtolength{\belowdisplayshortskip}{-2pt}
}

\fi

\usepackage{tikz}
\newcommand{\EmptyCircle}{%
  \tikz[baseline=-0.6ex]\draw[line width=0.4pt] (0,0) circle (0.8ex);%
}

\newcommand{\FullCircle}{%
  \tikz[baseline=-0.6ex]\fill (0,0) circle (0.8ex);%
}

\usepackage{makecell}

\usepackage[table]{xcolor}
\usepackage{color}

\usepackage{units}
\usepackage{soul}

\usepackage{caption}
\title{
    Criticality-Aware RAN Function Placement for Robust 6G Networks
}

\makeatletter
\newcommand{\linebreakand}{%
  \end{@IEEEauthorhalign}
  \hfill\mbox{}\par
  \mbox{}\hfill\begin{@IEEEauthorhalign}
}
\makeatother

\author{
\IEEEauthorblockN{Gabriel M. Almeida\IEEEauthorrefmark{1}, Joao F. Santos\IEEEauthorrefmark{2}, Walid Saad\IEEEauthorrefmark{3}, Nicola Marchetti\IEEEauthorrefmark{4}, and Kleber V. Cardoso\IEEEauthorrefmark{1}\\}

\IEEEauthorblockA{\IEEEauthorrefmark{1}Universidade Federal de Goiás, Brazil, E-mails: gabrielmatheus@inf.ufg.br, kleber@ufg.br\\
\IEEEauthorrefmark{2}Commonwealth Cyber Initiative (CCI), \IEEEauthorrefmark{3}Department of Electrical and Computer Engineering, Virginia Tech, USA, \\E-mails: \{joaosantos, walids\}@vt.edu\\
\IEEEauthorrefmark{4}Department of Electrical and Electronic Engineering, Trinity College Dublin, Ireland, E-Mail:  nicola.marchetti@tcd.ie\\
	}
}

\begin{document}
\bstctlcite{IEEEexample:BSTcontrol}

\makeatletter
\newcommand{\IEEEpubnotice}{%
  \vspace{-3em}
  \noindent
  \begin{minipage}{\textwidth}
    \footnotesize
    This work has been submitted to the IEEE for possible publication.\\
    Copyright may be transferred without notice, after which this version may no longer be accessible.
  \end{minipage}\par
  \vspace{0.5em}
  \hrule
  \vspace{0.51em}
}
\makeatother

\twocolumn[
\begin{@twocolumnfalse}
\IEEEpubnotice
\maketitle
\end{@twocolumnfalse}
]

\begin{abstract}
    The disaggregation of base stations into \ac{RAN} functions introduces new vulnerabilities to mobile networks, as disruptions in cloud locations can interrupt multiple \ac{RAN} functions and trigger cascading failures, leading to widespread outages.
    In this paper, we propose a criticality-aware \ac{RAN} function placement solution to improve the robustness of disaggregated mobile networks against random failures and targeted attacks.
    Our solution optimizes the placement of \acp{CU} and \acp{DU} across cloud locations to reduce functional dependencies and limit the criticality of cloud locations, while preserving the cost benefits of \ac{RAN} function centralization.
    We formulate the placement as a optimization model that jointly minimizes functional dependency and placement cost, subject to criticality constraints, as well as computing and transport requirements.
    We numerically evaluated our approach on a real-world mobile network topology across multiple disruption scenarios and demonstrated that our solution improves robustness, yielding up to 75\% more operational \acp{RU} after disruptions than existing approaches.
\end{abstract}
    

\acresetall    
   \section{Introduction}
\label{sec:intro}

One of the key features foreseen in the evolution toward 6G is the disaggregation of the \ac{RAN}, in which base stations are decomposed into \emph{function chains} of \acp{CU}, \acp{DU}, and \acp{RU}, each implementing different layers of the radio protocol stack~\cite{santos2025managing}.
This flexibility allows \acp{MNO} to place and scale \ac{RAN} functions across different \emph{cloud locations} according to computing resources, transport constraints, and traffic demand~\cite{Morais:2023}.
However, such flexibility has a cost, as disaggregation introduces new vulnerabilities to mobile networks~\cite{almeida:2026}.
For example, the disruption of a single cloud location can interrupt the function chains supporting the operation of multiple \acp{RU}, causing \emph{cascading failures}
across multiple cell sites~\cite{almeida:2026_2,Ferdowsi:21}.
In particular, the disruption of a \ac{DU} can interrupt service for all its associated \acp{RU}, while the disruption of a \ac{CU} can interrupt service for all its associated \acp{DU} and, subsequently, their associated \acp{RU}~\cite{almeida:2026}.

Disruptions in mobile networks can originate from multiple sources~\cite{Albert:2000}.
For example, \emph{random failures} can result from power outages, weather events, or equipment malfunctions, whereas \emph{targeted attacks} can deliberately target critical network components, potentially causing widespread outages by disrupting only a few cloud locations~\cite{almeida:2026}.
In both cases, the impact of a disruption depends on how \ac{RAN} functions are placed across cloud locations and their resulting \textit{functional dependencies}, which represent the dependence of network functions on the underlying infrastructure~\cite{Fang:23}, and \textit{criticality}, which reflects the importance of individual cloud locations to network operation and the consequences of their disruption~\cite{WalidSaad:2024}.
Consequently, the RAN function placement plays a key factor in network \emph{robustness} defined as the ability to withstand disturbances and maintain acceptable performance~\cite{WalidSaad:2024}.

Existing works on the robustness of mobile networks have largely focused on developing resource allocation mechanisms that accommodate variations in network demand and wireless channel conditions~\cite{Murti:23,Pires:2025}.
Only a few works investigate the robustness of disaggregated mobile networks against infrastructure disruptions, typically 
relying on introducing redundancies in the network, such as deploying backup \ac{RAN} functions, which increase computing resource consumption and deployment costs~\cite{Fang:23,Kang:22,Thiruvasagam:22,Kim:26}.
However, no existing approaches have investigated how the \ac{RAN} function placement can be leveraged to reduce the criticality of cloud locations, and consequently, improve the network's robustness against disruptions.

To address these limitations, the main contribution of this paper is to propose the first criticality-aware \ac{RAN} function placement solution to improve the robustness of disaggregated mobile networks to disruptions.
Our proposed solution: \textit{(i)} optimizes the placement of \ac{RAN} functions to reduce \acp{RU} functional dependency on cloud locations;
\textit{(ii)} improves network robustness against both random failures and targeted attacks while satisfying computing and transport requirements;
and \textit{(iii)} enables \acp{MNO} to balance robustness against disruptions with the cost savings achieved through RAN function centralization.
To the best of our knowledge, this is the first paper to explore RAN placement as a mechanism for reducing cloud location criticality and improving the robustness of disaggregated mobile networks. 
In summary, our key contributions include:
\begin{itemize}
    \item We propose the first criticality-aware \ac{RAN} function placement optimization model to improve the network's robustness against random failures and targeted attacks. 
    \item We show that network robustness can be improved through criticality-aware RAN function placement, without relying on additional functional redundancy.
    \item We validate our solution in a real-world mobile network topology, resulting in up to 75\% more operational \acp{RU} following disruptions compared to existing approaches.
\end{itemize}

The rest of this paper is organized as follows. In Section~\ref{sec:related_work}, we review background and related works. In Section~\ref{sec:system_model}, we describe our system model. In Section~\ref{sec:formulation}, we formulate our optimization problem. In Section~\ref{sec:evaluation}, we present numerical results validating our solution. Finally, in Section~\ref{sec:conclusion}, we pose our concluding remarks and offer directions for future work.
    \section{Related Works}
\label{sec:related_work}

In this section, we examine how functional dependencies between \acp{RAN} functions determine the criticality of cloud locations and influence network robustness.
Then, we compare existing approaches on the robustness of mobile networks. 

\subsection{Functional Dependency, Criticality, and Robustness}

In disaggregated mobile networks, providing mobile connectivity requires a functional chain comprising a \ac{CU}, a \ac{DU}, and an \ac{RU}, all of which must remain operational to serve users~\cite{Morais:2023}.
This requirement creates functional dependencies between \ac{RAN} functions as well as the cloud locations hosting them~\cite{Fang:23}.
Although each function chain comprises the same three functions, the placement of \acp{CU} and \acp{DU} across cloud locations determines the infrastructure components on which each physical \ac{RU} depends.
For example, the extent of these dependencies can be quantified by the number of cloud locations supporting each \ac{RU}, reflecting its exposure to single points of failure~\cite{Morais:2023,almeida:2026}.
These functional dependencies also determine the criticality of cloud locations, which reflects the potential consequences of their disruption on network operation.
For example, the criticality of a cloud location can be quantified by the number of \acp{RU} relying on the functions it hosts.

Consequently, the \textit{robustness of disaggregated mobile networks}, i.e., their ability to withstand disruptions in their infrastructure while maintaining acceptable operational performance~\cite{WalidSaad:2024}, is influenced by both the functional dependencies of \ac{RAN} functions and the criticality of cloud locations. For example, placing the \ac{CU} and \ac{DU} associated with an \ac{RU} across multiple cloud locations increases the number of cloud locations whose failure can interrupt the entire functional chain. Similarly, highly critical cloud locations can become attractive targets for adversaries, as compromising such locations may disrupt service to a large number of \acp{RU}.

\subsection{Existing Approaches on Network Robustness}

Most existing works focus on developing robust \ac{RAN} function placement approaches against traffic demand fluctuations under normal operating conditions, without considering how network operations would be affected by disruptions caused by compromised cloud locations.
These approaches often rely on adjusting functional split configurations, function placement, and computing resource allocation to accommodate demand fluctuations while satisfying computing and transport constraints.
For example, in~\cite{Morais:2023}, the authors proposed an optimal solution for both \ac{RAN} function placement and functional split selection, centralizing functions to reduce costs;
in~\cite{Murti:23}, the authors proposed a \ac{DRL}-based approach to dynamically optimize \ac{CU} and \ac{DU} placement under demand fluctuations.
Meanwhile, the work in~\cite{Pires:2025} proposed a framework that dynamically adapts \ac{CU} and \ac{DU} placement in response to traffic fluctuations, maximizing energy efficiency while satisfying computing and transport constraints.
However, these approaches~\cite{Morais:2023,Murti:23,Pires:2025} focus on adapting \ac{RAN} function placement to traffic fluctuations without considering the impact of disruptions on network operation.

The robustness of mobile networks against disruptions has received increasing attention in recent years~\cite{WalidSaad:2024}, with several works proposing approaches to mitigate the impact of disruptions to network infrastructure.
For example, in~\cite{Kang:22}, the authors proposed a robust function placement approach in the context of \ac{NFV}, considering the failures of network nodes, while the work in~\cite{Thiruvasagam:22} proposed a placement approach for service function chains that employs backup provisioning to satisfy reliability and delay requirements against random failures;
\cite{Kim:26} proposed an availability-aware placement and resource allocation approach that dynamically provisions dedicated or shared backup functions according to traffic demand, and~\cite{Fang:23}, the authors proposed a robust function placement approach that limits the number of users served by each function to mitigate the impact of failures and attacks.
Although these approaches~\cite{Kang:22,Thiruvasagam:22,Kim:26,Fang:23} improve robustness to disruptions, they rely on backup functions, which increase computing resource consumption and deployment costs.
Moreover, to the best of our knowledge, no existing works consider how the \ac{RAN} function placement determines the 
\acp{RU} functional dependencies to cloud locations, and the criticality of cloud locations, potentially creating shared points of failure that increase the impact of disruptions.
Table~\ref{tab:Related_work} compares existing approaches with our proposed solution.

\begin{table}[t]
\centering
\footnotesize
\setlength{\tabcolsep}{3pt}
\renewcommand{\arraystretch}{1.1}
\resizebox{\columnwidth}{!}{%
\begin{tabular}{|c|c|c|c|c|c|c|}
\hline
\rowcolor[HTML]{DAE8FC}
Ref. &
\begin{tabular}[c]{@{}c@{}}Function\\ Placement\end{tabular} &
\begin{tabular}[c]{@{}c@{}}Demand\\ Variation\end{tabular} &
\begin{tabular}[c]{@{}c@{}}Random\\ Failures\end{tabular} &
\begin{tabular}[c]{@{}c@{}}Targeted\\ Attacks\end{tabular} &
\begin{tabular}[c]{@{}c@{}}Functional\\ Dependency\end{tabular} &
Criticality \\ \hline
\cite{Morais:2023} &
\FullCircle & \EmptyCircle & \EmptyCircle &
\EmptyCircle & \EmptyCircle & \EmptyCircle \\

\rowcolor[HTML]{DAE8FC}
\cite{Murti:23} &
\FullCircle & \FullCircle & \EmptyCircle &
\EmptyCircle & \EmptyCircle & \EmptyCircle \\

\cite{Pires:2025} &
\FullCircle & \FullCircle & \EmptyCircle &
\EmptyCircle & \EmptyCircle & \EmptyCircle \\

\rowcolor[HTML]{DAE8FC}
\cite{Kang:22} &
\FullCircle & \EmptyCircle & \FullCircle &
\EmptyCircle & \EmptyCircle & \EmptyCircle \\

\cite{Thiruvasagam:22} &
\FullCircle & \EmptyCircle & \FullCircle &
\EmptyCircle & \EmptyCircle & \EmptyCircle \\

\rowcolor[HTML]{DAE8FC}
\cite{Kim:26} &
\FullCircle & \FullCircle & \FullCircle &
\EmptyCircle & \EmptyCircle & \EmptyCircle \\

\cite{Fang:23} &
\FullCircle & \FullCircle & \FullCircle &
\FullCircle & \EmptyCircle & \EmptyCircle \\ \hline

\rowcolor[HTML]{DAE8FC}
\textbf{This Work} &
\FullCircle & \FullCircle & \FullCircle &
\FullCircle & \FullCircle & \FullCircle \\ \hline

\end{tabular}%
}
\caption{Qualitative evaluation of related works.}
\label{tab:Related_work}
\vspace{-2em}
\end{table}

    \section{System Model}
\label{sec:system_model}

We consider a disaggregated mobile network that consists of cell sites hosting physical \acp{RU} and cloud locations capable of hosting \ac{DU} and \ac{CU}, e.g., edge, regional, and central clouds~\cite{santos2025managing}
These components are interconnected through a transport network, forming a distributed cloud computing architecture.
Each \ac{RU} is associated with a \ac{DU}, which in turn is associated to a \ac{CU} hosted at cloud locations, forming a function chain.
Finally, each function chain is connected to the \ac{CN} to provide mobile connectivity to users.

To capture the time-varying traffic demand of realistic networks, we consider a set of discrete time windows $\mathcal{T}$, where each $t \in \mathcal{T}$ represents an operational period.
Let $\mathcal{R}$ be the set of \acp{RU}, where each $r \in \mathcal{R}$ is characterized by its normalized traffic demand $\lambda_t(r) \in \mathbb{R}_{\geq 0}$ at time $t \in \mathcal{T}$.
We also consider a set $\mathcal{C}$ of cloud locations, where each $c \in \mathcal{C}$ has an available processing capacity at time $t \in \mathcal{T}$ captured by $\delta_t(c)$.
Since processing is the primary computing resource considered in \ac{RAN} function placement, we focus on the processing capacity of cloud locations and abstract away other resources~\cite{Morais:2023}.

The processing and transport requirements of \acp{CU} and \acp{DU} depend on the functional split configuration of the network.
In our model, we consider the 3GPP Option 2 functional split between \ac{CU} and \ac{DU}~\cite{3GPP:38801}, and the O-RAN 7.2x functional split between \ac{DU} and \ac{RU}~\cite{ORAN:WG4:CUS}.
Accordingly, \acp{CU} implements RRC and PDCP layers, while \acp{DU} implements RLC, MAC, and High-PHY layers.
Based on~\cite{Morais:2023}, we define $\Phi^{CU}$ and $\Phi^{DU}$ as the processing requirements of the \ac{CU} and \ac{DU} associated with an \ac{RU} operating at the maximum capacity, respectively.
Without loss of generality, we assume that processing requirements scale linearly with traffic demand~\cite{Rocha:2023}.
Then, the processing demands of the \ac{CU} and \ac{DU} associated with \ac{RU} $r \in \mathcal{R}$ at time $t \in \mathcal{T}$ are given by $\phi^{CU}_t(r) = \Phi^{CU} \lambda_t(r)$ and $\phi^{DU}_t(r) = \Phi^{DU} \lambda_t(r)$, respectively.

Based on~\cite{Morais:2023}, we denote by $\Theta^{CU}$, $\Theta^{DU}$, and $\Theta^{RU}$ the throughput requirements between the \ac{CN} and \acp{CU}, between \acp{CU} and \acp{DU}, and between \acp{DU} and \acp{RU}, respectively, which depend on the functional split and \ac{RAN} function implementation~\cite{Morais:2023}.
The throughput requirements associated with \ac{RU} $r \in \mathcal{R}$ at time $t \in \mathcal{T}$ are given by $\theta_t^{CU}(r) = \Theta^{CU}\lambda_t(r)$, $\theta_t^{DU}(r) = \Theta^{DU}\lambda_t(r)$, and $\theta_t^{RU}(r) = \Theta^{RU}\lambda_t(r)$, respectively.
Similarly, we define $\Delta^{CU}$, $\Delta^{DU}$, and $\Delta^{RU}$ as the maximum admissible transport latencies between the \ac{CN} and \acp{CU}, between \acp{CU} and \acp{DU}, and between \acp{DU} and \acp{RU}, respectively.

We represent the topology of the disaggregated mobile network and the cloud computing infrastructure as a graph $\mathcal{G} = (\mathcal{V}, \mathcal{E})$, where $\mathcal{V} = \mathcal{R} \cup \mathcal{C} \cup \{v_0\}$ represents the set of nodes, composed of the \acp{RU}, the cloud locations, and the \ac{CN} denoted by $v_0$.
The set of links represents the transport network infrastructure, and is given by $\mathcal{E} = \{(i,j); i, j \in \mathcal{V}\}$, where each link from $i$ to $j$ is represented by the tuple $(i,j)$ with an available transport capacity $\gamma_t(i,j)$ at time $t$.
We consider a set of paths $\mathcal{P}_{m,n} $ between $m, n \in \mathcal{V}$ as a sequence of links $(i, j)$, where $i, j \in \mathcal{V}$.
$I_p(i, j)$ indicates if link $(i, j)$ is in path $p \in \mathcal{P}_{m,n}$.
Finally, $\omega_t(p)$ represents the end-to-end transport latency along path $p \in \mathcal{P}_{m,n}$ at time $t \in \mathcal{T}$.
    \section{Criticality-Aware RAN Function Placement}
\label{sec:formulation}

In this section, we present the mathematical formulation of our criticality-aware \ac{RAN} function placement optimization model.
Our goal is to improve the robustness of disaggregated mobile networks against random failures and targeted attacks by optimizing \ac{RAN} function placement while accounting for the functional dependency of \acp{RU} and the criticality of cloud locations.
To model the \ac{RAN} function placement, we define two binary decision variables, $x_t^{c,r}, y_t^{c,r} \in \{0,1\}$, indicating if cloud location $c \in \mathcal{C}$ hosts the \ac{CU} or \ac{DU} associated with \ac{RU} $r \in \mathcal{R}$ at time $t \in \mathcal{T}$, respectively.
Similarly, we define three binary decision variables, $f_t^{p,r}, g_t^{p,r}, h_t^{p,r} \in \{0,1\}$, to indicate whether path $p \in \mathcal{P}_{m,n}$ is selected to route traffic between the \ac{CN} and \ac{CU}, \ac{CU} and \ac{DU}, and \ac{DU} and \ac{RU}, respectively.

We use $z_t^{c,r} \in \{0,1\}$ to indicate whether cloud location $c \in \mathcal{C}$ hosts the \ac{CU} or \ac{DU} function of \ac{RU} $r \in \mathcal{R}$ at time $t \in \mathcal{T}$.
We formulate the total functional dependency of \acp{RU}, representing the number of cloud locations that their function chains depend on, at time $t \in \mathcal{T}$ as:
\begin{equation}
D_t = \sum_{r \in \mathcal{R}} \sum_{c \in \mathcal{C}} z_t^{c,r}.
\label{eq}
\end{equation}

To capture the criticality of cloud locations, representing the amount of \acp{RU} whose function chains depend on a given cloud location, we formulate the criticality of a cloud location $c \in \mathcal{C}$ at time $t \in \mathcal{T}$ as:
\begin{equation}
K_t(c) = \frac{1}{|\mathcal{R}|} \sum_{r \in \mathcal{R}} z_t^{c,r},
\qquad \forall c \in \mathcal{C}.
\label{eq}
\end{equation}

To capture placement cost, we consider \ac{RAN} function centralization and the active cloud locations, following~\cite{Morais:2023}.
Higher function centralization and fewer active cloud locations reduce the computational resources required to operate the network, lowering placement costs.
Let $w_t^c \in \{0,1\}$ indicate if cloud location $c \in \mathcal{C}$ hosts at least one \ac{RAN} function.
We formulate the placement cost as:
\begin{equation}
O_t = \sum_{c \in \mathcal{C}} w_t^c -
\sum_{c \in \mathcal{C}}\sum_{r \in \mathcal{R}}
(x_t^{c,r}+y_t^{c,r}).
\label{eq}
\end{equation}

Our criticality-aware \ac{RAN} function placement problem involves the joint minimization of the \acp{RU} functional dependency, the criticality of cloud locations, and cost.
Minimizing functional dependency favors placing the \acp{CU} and \acp{DU} of the same function chain in the same cloud location, reducing the number of cloud locations it depends on.
Similarly, minimizing placement cost favors centralizing \ac{RAN} functions in fewer cloud locations, reducing the number of active cloud locations.
In contrast, minimizing the criticality of cloud locations favors distributing \ac{RAN} functions across multiple cloud locations, limiting the number of \acp{RU} that depend on the same cloud location and reducing their criticality.
Therefore, these objectives introduce a trade-off between centralizing \ac{RAN} functions to reduce dependencies and costs and distributing them to limit the impact of disruptions at individual cloud locations.

To address this tradeoff, the challenge is to limit cloud location criticality while minimizing functional dependency and placement cost.
Although these objectives could be combined using a weighted-sum formulation, selecting appropriate weights is challenging due to their different scales and competing effects.
Instead, we adopt an $\varepsilon$-constraint approach~\cite{Fang:23}, which imposes an upper bound on the criticality of cloud locations while jointly minimizing the total functional dependency and placement cost.
This parameterization enables \acp{MNO} to specify the maximum acceptable fraction of \acp{RU} for any individual cloud location and to balance deployment cost tradeoffs.
We formulate the resulting optimization problem, solving it independently for each $t \in \mathcal{T}$, as:
\begin{equation}
    \text{P1}: \quad \minimize_{x_t^{c,r}, y_t^{c,r}, f_t^{p,r}, g_t^{p,r}, h_t^{p,r}} \sum_{t \in \mathcal{T}}
    \bigg ( D_t + O_t \bigg ).
\label{eq}
\end{equation}
subject to:
\begin{gather}
   \quad K_t(c) \leq \varepsilon,
    \qquad \forall c \in \mathcal{C}, t \in \mathcal{T},
    \label{eq}
\end{gather}
where $\varepsilon \in [0,1]$ limits the criticality of individual cloud locations, bounding the fraction of \acp{RU} potentially disrupted by a single cloud location failure.
Our optimization model is also subject to the following constraints.

To ensure that the function chain associated with each \ac{RU} has at most one \ac{DU} and one \ac{CU} at time $t \in T$:
\begin{equation}
    \sum_{c \in \mathcal{C}} x_t^{c, r} \leq 1, \qquad \forall r \in \mathcal{R}, t \in \mathcal{T},
    \label{eq:at_most_one_CU}
\end{equation}
\begin{equation}
    \sum_{c \in \mathcal{C}} y_t^{c, r} \leq 1, \qquad \forall r \in \mathcal{R}, t \in \mathcal{T}.
    \label{eq:at_most_one_DU}
\end{equation}

Each cloud location can host multiple \acp{CU} and \acp{DU}.
To ensure that the total processing demand of \acp{CU} and \acp{DU} in each cloud location $c \in \mathcal{C}$ does not exceed its available computational capacity at time $t \in \mathcal{T}$, we 
impose the following constraint:
\begin{equation}
    \sum_{r \in \mathcal{R}} \Bigg ( x_t^{c, r} \phi_t^{CU}(r) + y_t^{c, r} \phi_t^{DU}(r)\Bigg ) \leq \delta_t(c), \forall c \in \mathcal{C}, t \in \mathcal{T}.
    \label{eq:computational_capacity}
\end{equation}

To ensure connectivity between the \ac{RAN} functions in each function chain, a routing path $p \in \mathcal{P}_{m,n}$ must be established 
for each transport link
between \ac{CN} and \ac{CU}, between \ac{CU} and \ac{DU}, and between \ac{DU} and \ac{RU}.
We enforce this by the following constraints:
\begin{equation}
    \sum_{p \in \mathcal{P}_{n_0,n}} x_r^{n, r} f_t^{p, r} = 1,
    \label{eq:routing_1}
\end{equation}
\begin{equation}
    \sum_{p \in \mathcal{P}_{m,n}} x_r^{m, r} g_t^{p, r} y_r^{c, n} = 1,
    \label{eq:routing_2}
\end{equation}
\begin{equation}
    \sum_{p \in \mathcal{P}_{m,r}} y_r^{c, m} h_t^{p, r} = 1,
    \label{eq:routing_3}
\end{equation}
\noindent where the nonlinear products in~\eqref{eq:routing_1} -- \eqref{eq:routing_3} can be linearized using standard techniques for binary variables, following~\cite{Glover:1975}.

The capacity constraints of the transport links $(i, j) \in \mathcal{E}$ must be satisfied to ensure that the total traffic routed through each link does not exceed available transport capacity at time $t \in \mathcal{T}$.
We express this requirement as follows:
\begin{align}
    \sum_{r \in \mathcal{R}} &\sum_{p \in \mathcal{P}_{m,n}} \Bigg ( I_p(i, j) \bigg [ f_t^{c, r} \theta_t^{CU}(r) + g_t^{c, r} \theta_t^{DU}(r) \\ \noindent &+ h_t^{c, r} \theta_t^{RU}(r) \bigg ] \Bigg) \leq \gamma_t(i, j), \qquad \forall(i, j) \in \mathcal{E}, t \in \mathcal{T}.
    \label{eq:transport_capacity}
\end{align}

The latency of the selected routing paths $p \in \mathcal{P}_{m,n}$ between the \ac{RAN} functions in each of the function chains must not exceed the maximum latency requirement of the transport links between \ac{CN} and \ac{CU}, between \ac{CU} and \ac{DU}, and between \ac{DU} and \ac{RU}. 
We express this requirement as follows:
\begin{equation}
    \sum_{p \in \mathcal{P}_{m,n} f_t^{p, r}} \omega_t(p) \leq \Delta^{CU}, \qquad \forall r \in \mathcal{R}, t \in \mathcal{T},
    \label{eq:latency_1}
\end{equation}
\begin{equation}
    \sum_{p \in \mathcal{P}_{m,n} g_t^{p, r}} \omega_t(p) \leq \Delta^{DU}, \qquad \forall r \in \mathcal{R}, t \in \mathcal{T},
    \label{eq:latency_2}
\end{equation}
\begin{equation}
    \sum_{p \in \mathcal{P}_{m,n} h_t^{p, r}} \omega_t(p) \leq \Delta^{RU}, \qquad \forall r \in \mathcal{R}, t \in \mathcal{T}.
    \label{eq:latency_3}
\end{equation}

The P1 objective together with the above constraints, defines a \ac{MINLP} problem due to the products of binary decision variables in Equations \eqref{eq:routing_1} -- \eqref{eq:routing_3}.
These products can be linearized, resulting in an equivalent \ac{MILP} formulation, which is NP-hard~\cite{Morais:2023}.
Nevertheless, as shown in the next section, we obtain optimal solutions for relevant instances derived from real-world network topologies across multiple disruption scenarios using CPLEX.
    \section{Simulation Results and Analysis}
\label{sec:evaluation}

In this section, we evaluate our criticality-aware \ac{RAN} function placement solution 
to improve the network's robustness against disruptions.
First, we describe our simulation environment and the disruption scenarios.
Then, we assess the robustness of our proposed solution, investigate the impact of the criticality threshold, and evaluate resource utilization.

\subsection{Evaluation Scenario}

\begin{figure}[t]
    \centering
    \begin{subfigure}[b]{0.49\columnwidth}
        \centering
        \includegraphics[width=0.90\textwidth]{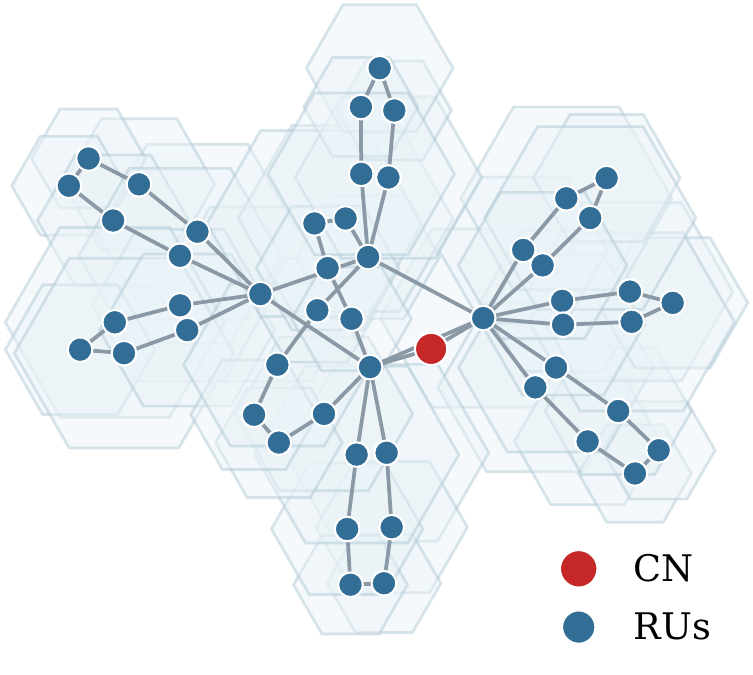}
        \vspace{-1em}
        \caption{Real mobile network topology comprising 50 \acp{RU} and the \ac{CN}~\cite{5Gxcrosshaul:17}.}
        \label{fig:topology}
    \end{subfigure}
    \hfill
    \begin{subfigure}[b]{0.49\columnwidth}
        \centering
        \includegraphics[width=\textwidth]{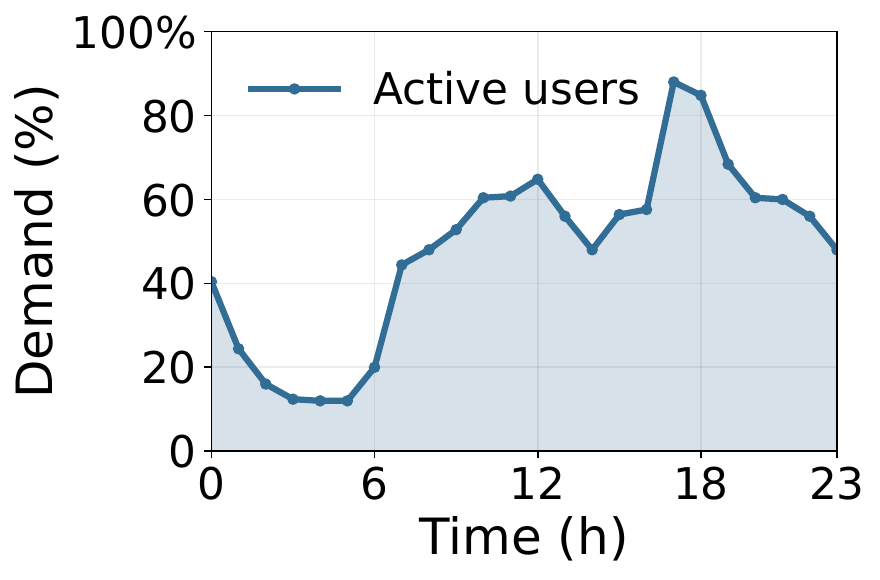}
        \vspace{-2em}
        \caption{Traffic demand variation based on a profile of active users~\cite{Pires:2025}.}
        \label{fig:demand_variation}
    \end{subfigure}
    \vspace{0.1em}
    \caption{Scenario in our numerical evaluations, comprising a mobile network topology and traffic demand variation.}
    \label{fig:demand_variation_topology}
    \vspace{-1em}
\end{figure}

We developed a simulator to numerically evaluate our solution using a real-world mobile network topology from the \textit{5G-Crosshaul} project~\cite{5Gxcrosshaul:17}.
The topology comprises 50 \acp{RU}, each 
co-located with a cloud location, and configured with \unit[100]{MHz} bandwidth, 32 antenna ports, 8 MIMO layers, and 256-QAM modulation. 
All cloud locations are interconnected through a transport network, as illustrated in Fig.~\ref{fig:topology}.
The network serves a total of 6000 users based on the user density reported in~\cite{Esper:2023}.
To capture temporal traffic fluctuations, we used the simulator from~\cite{Pires:2025} to generate 
an hourly network-wide traffic demand profile, expressed as a percentage of active users at each hour, over seven days (168 hours), as shown in Fig.~\ref{fig:demand_variation}.

We evaluate network robustness under two disruption models: random failures and targeted attacks.
For random failures, we generate 100 independent realizations of disrupted cloud locations using different random seeds, assuming equal failure probabilities across all cloud locations.
For targeted attacks, cloud locations are progressively disrupted in descending order of \textit{betweenness centrality}, a graph-theoretical metric that quantifies a node's importance based on the fraction of shortest paths between other nodes that traverse it~\cite{barabasi:2016}.

We compare our solution against three benchmarks:
\emph{(i) D-RAN}, a fully distributed architecture in which all \ac{RAN} functions are deployed at the cell sites hosting their respective \acp{RU};
\emph{(ii) PlaceRAN}~\cite{Morais:2023}, a state-of-the-art \ac{RAN} function placement solution that prioritizes function centralization while minimizing costs; and \emph{(iii) Reveal}~\cite{Fang:23}, a robust virtualized function placement solution that minimizes deployment costs while restricting the number of users dependent on individual functions.
To allow comparison with~\cite{Fang:23}, we adopt a criticality threshold of $\varepsilon = 0.1$, limiting the dependency on any cloud location to at most 10\% of \acp{RU}.
We acknowledge that the optimal criticality threshold may vary with topology and user distribution, and we plan to investigate this in future work.

\subsection{Functional Dependency and Robustness Against Failures}

\begin{figure}[t]
    \centering
    \includegraphics[width=0.95\linewidth]{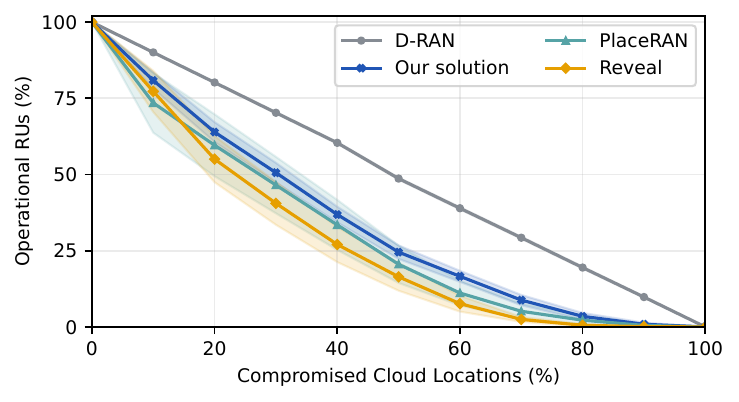}
    \vspace{-1em}
    \caption{Operational \acp{RU} after random failures:  our solutions results in up to 20\% higher robustness than existing approaches.}
    \label{fig:robustness_failrues}
    \vspace{-0.8em}
\end{figure}

\begin{figure}[t]
    \centering
    \includegraphics[width=0.95\linewidth]{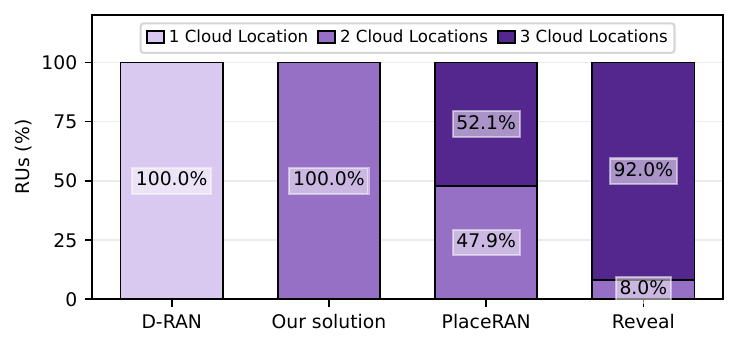}
    \vspace{-1em}
    \caption{Comparison of \acp{RU} functional dependencies: we achieve higher robustness due to lower functional dependency.}
    \label{fig:functional_dependency}
    \vspace{-1em}
\end{figure}

First, we are interested in evaluating the robustness of our solution against random failures.
Figure~\ref{fig:robustness_failrues} shows the percentage of operational \acp{RU} as cloud locations are progressively compromised by random failures.
Our solution results in up to 12\% and 18\% more operational \acp{RU}
than PlaceRAN and Reveal, respectively.
In addition, we can observe that our solution results in a robustness very close to the D-RAN baseline, remaining, on average, only 10\% below its performance.
However, the D-RAN achieves the highest robustness at the cost of sacrificing the cost savings of \ac{RAN} function centralization, as discussed further in Section~\ref{subsec:centralization_and_cloud_location_activation}. 
To understand how functional dependencies contribute to these results, Fig.~\ref{fig:functional_dependency} presents the distribution of \acp{RU} according to their functional dependencies.
For PlaceRAN and Reveal, 52.1\% and 92\% of \acp{RU} depend on three cloud locations, respectively, whereas our solution ensures that all \acp{RU} depend on two cloud locations.
In contrast, D-RAN results in all \acp{RU} depend on a single cloud location, as all \ac{RAN} functions are collocated at their respective cell sites.
These results demonstrate that distributing \ac{RAN} functions across more cloud locations increases the network's vulnerability to failures by introducing additional potential failure points for each function chain.
By reducing these functional dependencies, our RAN function placement solution improves network robustness against random failures,
 limiting the number of cloud locations whose disruption can interrupt function chains and the operation of \acp{RU}.

\subsection{Criticality and Robustness Against Attacks}

\begin{figure}[t]
    \centering
    \includegraphics[width=0.95\linewidth]{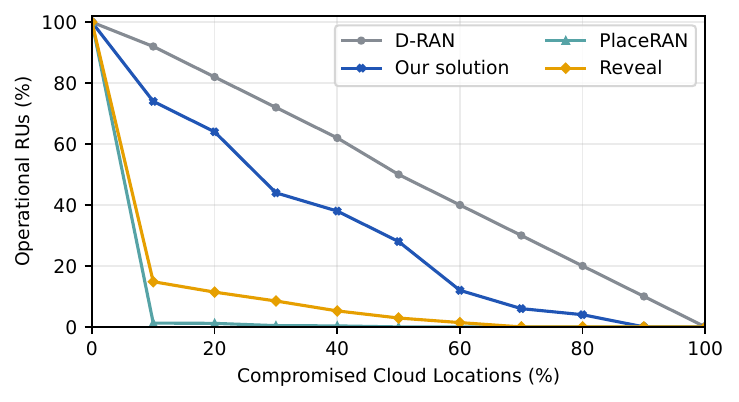}
    \vspace{-1em}
    \caption{Operational \acp{RU} after targeted attacks: our solution results in up to 75\% higher robustness than existing approaches.}
    \vspace{-0.8em}
    \label{fig:robustness_attacks}
\end{figure}

\begin{figure}[t]
    \centering
    \includegraphics[width=0.95\linewidth]{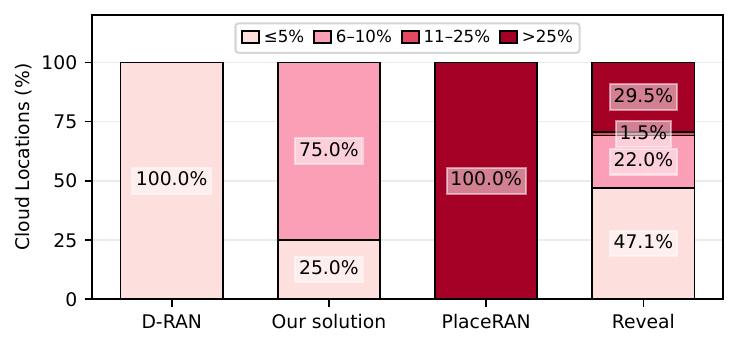}
    \vspace{-1em}
    \caption{Comparison of the criticality of cloud locations: we achieve higher robustness by enforcing a lower criticality.}
    \vspace{-1em}
    \label{fig:criticality}
\end{figure}

Next, we are interested in evaluating the robustness of our solution against targeted attacks.
Figure~\ref{fig:robustness_attacks} shows the percentage of operational \acp{RU} as cloud locations are progressively disrupted by targeted attacks.
Our solution results in up to 60\% and 75\% more operational \acp{RU} than PlaceRAN and Reveal, respectively.
We can observe that our solution performs significantly better against targeted attacks, remaining, on average, only 20\% below the D-RAN baseline.
To understand how cloud location criticality contributes to these results, Fig.~\ref{fig:criticality} shows the distribution of cloud locations by criticality.
For PlaceRAN and Reveal, 100\% and 29.5\% of cloud locations exhibit criticality above 25\%, respectively, while our solution results in 75\% of cloud locations remaining within the 6--10\% criticality range.
In contrast, D-RAN results in all cloud locations remaining below 5\% criticality.
These results demonstrate that our solution improves the robustness of mobile networks against targeted attacks by limiting the criticality of cloud locations, reducing the number of \acp{RU} affected by disruptions.

\subsection{Centralization and Cloud Location Activation}
\label{subsec:centralization_and_cloud_location_activation}

\begin{figure}[t]
    \centering
    \hfill
    \begin{subfigure}[b]{0.42\columnwidth}
        \centering
        \includegraphics[width=\textwidth]{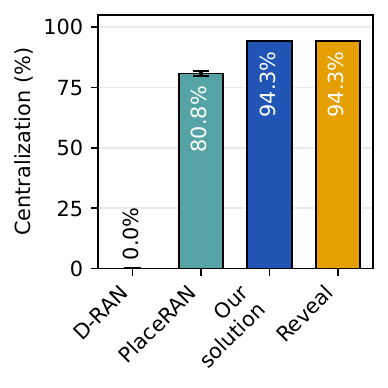}
        \vspace{-2em}
        \caption{Centralization.}
        \label{fig:centralization}
    \end{subfigure}
    \hfill
    \begin{subfigure}[b]{0.42\columnwidth}
        \centering
        \includegraphics[width=\textwidth]{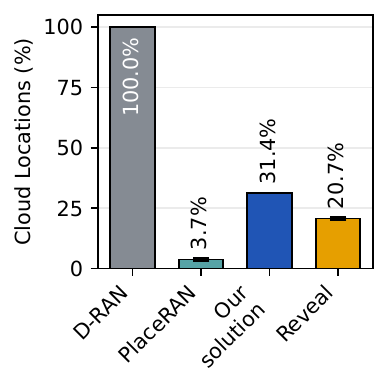}
        \vspace{-2em}
        \caption{Used Cloud Locations.}
        \label{fig:used_cloud_locations}
    \end{subfigure}
    \hfill
    \vspace{0.5em}

    \caption{Centralization and active cloud locations: our gains in robustness require only 10.7\% more active cloud locations.}
    \label{fig:centralization_used_cloud_locations}
    \vspace{-1em}
\end{figure}

In this analysis, we assess the deployment costs associated with the different \ac{RAN} placement approaches, in terms of their \ac{RAN} function centralization and the number of active cloud locations required by each solution. 
This is essential to understanding the tradeoff between network robustness and the additional computational resources needed to support the network infrastructure, as reducing the criticality of cloud locations and the functional dependency of function chains may require activating additional cloud locations.
Figure~\ref{fig:centralization} shows the degree of centralization, and Fig.~\ref{fig:used_cloud_locations} shows the percentage of active cloud locations for each approach.
By definition, D-RAN does not centralize RAN functions and requires all cloud locations to remain active, resulting in higher deployment costs~\cite{Morais:2023}. 
In contrast, PlaceRAN prioritizes minimizing the number of active cloud locations without considering robustness.
Our solution and Reveal achieve higher degrees of centralization than PlaceRAN, although they require more active cloud locations.
Compared with Reveal, our solution achieves the same degree of centralization while activating only 10.7\% more cloud locations.
These results demonstrate that our solution improves network robustness  against random failures and targeted attacks without sacrificing the benefits of \ac{RAN} function centralization, 
They also highlight the tradeoff between network robustness and deployment costs. In particular, our solution achieves significantly higher robustness than PlaceRAN and Reveal while requiring fewer active cloud locations than Reveal but more than PlaceRAN.
    \section{Conclusion}
\label{sec:conclusion}

In this paper, we have introduced a criticality-aware \ac{RAN} function placement solution to improve the robustness of disaggregated mobile networks against random failures and targeted attacks.
Our solution minimizes the functional dependencies of \acp{RU} and limits the criticality of cloud locations, reducing the impact of disruptions while preserving the benefits of \ac{RAN} function centralization.
We have evaluated our solution on a real-world mobile network topology, demonstrating its ability to maintain more operational \acp{RU} after disruptions and achieving up to 75\% greater robustness than existing placement approaches.
Our results also highlight the trade-off between robustness against random failures and targeted attacks, \ac{RAN} function centralization, and placement cost, demonstrating that improving robustness does not necessarily require sacrificing function centralization, although it may require activating additional cloud locations.
In future work, we plan to \textit{(i)} assess the impact of different values of $\varepsilon$ on network robustness and investigate strategies for determining its optimal value, 
\textit{(ii)} investigate how the \ac{RAN} function placement before disruptions influences network recovery, and \textit{(iii)} explore the impact of transport network failures and the criticality of transport links.

\section*{Acknowledgments}
The research leading to this paper received support from the Commonwealth Cyber Initiative.
For more information, visit: \url{www.cyberinitiative.org}.
This work also received support from the National Science Foundation US-Ireland R\&D Partnership program under grant No. 2421362 (Resilient Networks project).
This work was supported by CAPES, by MCTIC/CGI.br/FAPESP under grant no. 2020/05127-2, by CNPq under grant no. 306283/2025-5, by RNP/MCTIC under grant no. 01245.020548/2021-07, and by the OpenRAN Brazil project under grant A01245.014203/2021-14.
    
    \printbibliography

\end{document}